\documentclass[doublecol]{epl2} 

\usepackage{graphicx}
\usepackage{dcolumn}
\usepackage{bm}
\usepackage{amssymb}
\usepackage{hyperref}
\usepackage{multirow}
\usepackage{color}
\usepackage{amsmath}

\newcommand{\Da}{\mathcal{D}}

\title{Velocity alignment promotes motility-induced phase separation}

\author{E. Ses\'{e}-Sansa\inst{1} \and I. Pagonabarraga\inst{1,2,3} \and D. Levis\inst{1,2,3}}
\shortauthor{E. Ses\'{e}-Sansa, I. Pagonabarraga, D. Levis}

\institute{   
  \inst{1} CECAM Centre Europ\' een de Calcul Atomique et Mol\'eculaire, \'Ecole Polytechnique F\'ed\'erale de Lausanne, Batochimie, Avenue Forel 2, 1015 Lausanne, Switzerland \\                
  \inst{2} Departament de F\'{\i}sica de la Mat\`{e}ria Condensada, Universitat de Barcelona, Mart\'{\i} i Franqu\'{e}s 1, E08028 Barcelona, Spain\\
  \inst{3} UBICS University of Barcelona Institute of Complex Systems, Mart\'i i Franqu\`es 1, E08028 Barcelona, Spain\\
}

\abstract{
We study the phase behavior of polar Active Brownian Particles moving in two-spatial dimensions and  interacting through volume exclusion and  velocity alignment. We combine particle-based simulations of the microscopic model with a simple mean-field kinetic model to understand the impact of velocity alignment on the motility-induced phase separation of  self-propelled disks. We show that, as the alignment strength is increased, approaching the onset of collective motion from below, orientational correlations grow, rendering the diffusive reorientation dynamics slower. As a consequence,  the tendency of particles to aggregate into isotropic clusters is enhanced,  favoring the complete de-mixing of the system into a low and high-density phase. 
}

\begin{document}

\maketitle

Active systems made of  self-propelled particles have been (and still are) the object of intense research 
 \cite{RamaswamyRev, VicsekRev,MarchettiRev}. Such an interest is mainly due to the fact that, unlike passive systems
 , active systems are driven out-of-equilibrium at the level of a single constituent, giving rise to a rich and novel phenomenology. 
In particular, activity may trigger non-equilibrium large-scale behavior like clustering and phase separation in the absence of cohesive forces, or the spontaneous emergence of collective directed motion.
For instance, bacterial colonies self-organize into coherently moving swarms \cite{Zhang2010, Peruani2012,Chate2017bac}, self-propelled Janus colloids form dynamical clusters \cite{Theurkauff2012, Palacci2013,  Buttinoni2013, BechingerRev, Ginot2015, Ginot2018} and actin filaments self-organize into dense traveling structures at high density \cite{Schaller2010,Huber2018}. 
Similarly, large-scale directed motion has been observed in monolayers of polar grains and colloidal rollers suspensions  \cite{Deseigne2010, Bricard2013}.      

In order to gain understanding on the general mechanisms underlying such collective phenomena, several simplified model systems have been introduced and analyzed in depth. 
In particular, the role played by the nature (or symmetry) of the microscopic interactions between agents on  phase transitions and structure formation has been extensively discussed.  
Among these minimal models, the Active Brownian Particle (ABP)  and Vicsek model, together with their variants, occupy a privileged place. The ABP  model  \cite{Romanczuk2012active, Fily2012, Redner2013, CatesTailleur2013, Bialke2013,  Stenhammar2014, Levis2017, Digregorio2018} introduces the simplest interaction mechanism between isotropic particles, i.e. volume exclusion, while the Vicsek model \cite{Vicsek1995, Toner1995, Gregoire2004, Chate2008, Chate2008Rev, Mishra2010, Solon2015, VicsekRev} considers a local 'ferromagnetic' velocity alignment mechanisms between otherwise non-interacting polar agents. As such, they constitute the prototypical, and therefore most studied, models of isotropic and polar active matter. 
In the ABP model, the tendency to form dense clusters arises from the 'mere' combination of self-propulsion and excluded volume interactions. These clusters grow as self-propulsion or density is increased and eventually lead to a complete phase separation, the so-called Motility-Induced phase separation (MIPS) \cite{CatesRev} originally reported in the context of run-and-tumble particles \cite{TailleurCates2008}.  
 In the Vicsek model, the direction of self-propulsion of a particle tends to align towards the mean direction of its neighbors. Such an alignment  interaction triggers an order-disorder phase transition \cite{Vicsek1995, Toner1995, VicsekRev} characterized by the emergence of different kinds of traveling structures \cite{Gregoire2004, Chate2008, Chate2008Rev, Mishra2010}. 

In experimental set-ups, interactions between components obviously involve more ingredients than the simplified ones captured  by the ABP or Vicsek models.
Therefore, the impact of the interplay between these two paradigmatic interactions, excluded volume and velocity alignment,  on the large-scale behavior of active systems appears as a natural question which has motivated the introduction of several models \cite{Szabo2006, Henkes2011, Peruani2011traffic, Farrell, Lam2015, PeruaniChetrite, Aitor}.
However, such a question has been mainly addressed from continuum models where excluded volume effects are captured through an effective coarse-grained density-dependent self-propulsion velocity \cite{Farrell, PeruaniChetrite}. 
While this mesoscopic description is well-grounded for isotropic active matter, for which it can be systematically derived from the microscopic dynamics \cite{Tailleur2008, CatesTailleur2013, Bialke2013, SharmaBrader}, such a connection between the microscopic and mesoscopic description is still lacking in the presence of alignment. 

Here, in order to clarify the role played by alignment on the phase behavior of ABP, we keep a microscopic description of the interactions. 
We show that, below the onset of  polar order (or flocking),  the self-trapping mechanism responsible of MIPS is maintained in the presence of velocity-alignment. We show that alignment favors the emergence of MIPS, in agreement with the continuum formulation in \cite{PeruaniChetrite}. Our simulations, complemented  by a simple kinetic argument based on \cite{Redner2013}, allow us to understand the microscopic origin of such behavior: alignment induces orientational correlations which slow down the random reorientation of the particles, thus increasing the persistence of their motion.

\section{Model and methods}
We study a two dimensional system of $N$ self-propelled particles in a $L\times L$ box with periodic boundary conditions \cite{Levis2018JCMP}. Each particle at position $\textbf{r}_{i}(t) = (x_{i},y_{i})$ is self-propelled with constant speed $v_{0}$ along the direction given by $\textbf{n}_{i}=(\cos\theta_{i},\sin\theta_{i})$. 
The over-damped Langevin equations governing the evolution of the system are
\begin{equation}
\dot{\textbf{r}}_{i} = v{_0}\textbf{n}_{i} + \mu \textbf{F}_{i}+ \sqrt{2D_{0}}\boldsymbol{\eta}_{i}
\label{eq:eomR}
\end{equation}
\begin{equation}
\dot{\theta}_{i} = \frac{K}{ \pi R^{2}}\sum_{j\in \omega_{i}}\sin(\theta_{j}-\theta_{i}) + \sqrt{2D_{\theta}}\nu_{i} 
\label{eq:eomT}
\end{equation}
where $\boldsymbol{\eta}_{i}$ and $\nu_{i} $ are two independent white Gaussian noises with zero mean and unit variance: $\langle {\eta}_{i}^{\alpha}(t) {\eta}^{\beta}_{j}(t^{\prime}) \rangle=\delta_{ij}\delta^{\alpha \beta}\delta(t-t^{\prime})$ and  $\langle {\nu}_{i}(t) {\nu}_{j}(t^{\prime})\rangle=\delta_{ij}\delta(t-t^{\prime})$ ($\alpha$, $\beta$ denoting cartesian coordinates). The noise $\boldsymbol{\eta}_{i}$ represents a thermal bath at temperature $T$. The diffusivity $D_0$ and mobility $\mu$  verify $D_0=\mu k_B T$.
Particles are subjected to both excluded volume and velocity alignment interactions. The inter-particle force, $\boldsymbol{F}_{i}=-\sum_{j\neq i}\nabla_i U(r_{ij})$,  accounts for short-range repulsions which we model by a repulsive potential of the form $U(r) = u_{0}({\sigma}/{r})^{12}$, 
with an upper cut-off at $3 \sigma$. 
Velocity alignment is introduced as a torque in terms of the sum of the phase difference and its strength is controlled by the 'ferromagnetic' coupling parameter $K\geq 0$ \cite{Gregoire2004,Peruani2008mean, Grossmann2012, PeruaniChetrite, Farrell, Aitor,  Levis2018JCMP}. 
The sum in eq. \ref{eq:eomT} runs over the particles in the vicinity of particle \textit{i}, denoted $\omega_{i}$, defined by the interaction range, $R$, which is chosen to be slightly larger than $\sigma$. 
This mimics an alignment mechanism which is of non-steric origin, meaning that  particles do not need to be in contact in order to align, as in  \cite{Szabo2006, Henkes2011,Lam2015}. As such, it is closer to the alignment mechanism of flocking birds or hydrodynamically coupled micro-swimmers \cite{VicsekRev, Bricard2013,Alarcon2013}.   

It is convenient at this stage to identify the relevant set of dimensionless parameters.
The parameters  $\sigma$, $\tau = D_{\theta}^{-1}$ and $u_{0}$ provide the natural units of  length, time and energy  of the model, respectively. Here, we fix $R=2\sigma$ and $D_{\theta} =3 D_{0}/\sigma^{2}$. Then, we are left with the reduced coupling parameter $g = \frac{K}{4\pi \sigma^{2} D_{\theta} }$ quantifying the strength of the alignment interaction with respect to angular diffusion; the P\'{e}clet number $\text{Pe} = \frac{v_{0}}{\sigma D_{\theta}}$ and the mean density $\phi=\rho\pi/4$ (where $\rho=N/L^2$). The  ratio $\Gamma=\frac{u_{0} \mu}{\sigma v_{0} }$ accounting for the competition between the softness of the potential and the self-propulsion strength should, in general, be considered as well. However, in the present study we focus on a stiff potential which diverges at contact and in a regime of (small enough) $v_0$ which ensures that repulsive forces always dominate over self-propulsion at short distances. 

For $g=0$ our model reduces to the paradigmatic ABP model, for which it is known that at high enough density and self-propulsion velocity it undergoes  MIPS   \cite{Fily2012, Redner2013, Bialke2013, Stenhammar2014, Levis2017, Digregorio2018}. For $u_0=0$, in turn, the model is equivalent to a variant of the Vicsek model in continuous time \cite{Grossmann2012, Farrell, PeruaniChetrite, Levis2018JCMP}, which exhibits a transition towards collective motion above a critical value of the coupling $g$. Thus, our model includes the main ingredients of two paradigmatic models of active matter, introducing a direct coupling between the velocity and the density of aligning active particles. 

Here, we numerically solve the equations of motion \ref{eq:eomR} and \ref{eq:eomT} using a Euler integrator with time step $\Delta t = 10^{-3}$.  
We simulate  systems of $N=4000$ particles at fixed $\phi=0.4$,  $D_{\theta} = 0.005$ 
 and $\mu=1$. 
We explore the phase behavior of the system in the $g-\text{Pe}$ plane by varying $v_0$ from $0$ to $0.3$ (thus implying $\text{Pe}\in[0:60]$)  
 and $K$ from $0$ to $0.6$ (thus $g \in [0, 5]$).
To analyze the long-time behavior of the system, we let it evolve from a homogeneous initial state until it reaches  stationarity. The measurements reported have been obtained by averaging over $10^4$ independent steady-state configurations.

\section{Flocking}
\begin{figure}
\onefigure[scale=0.65]{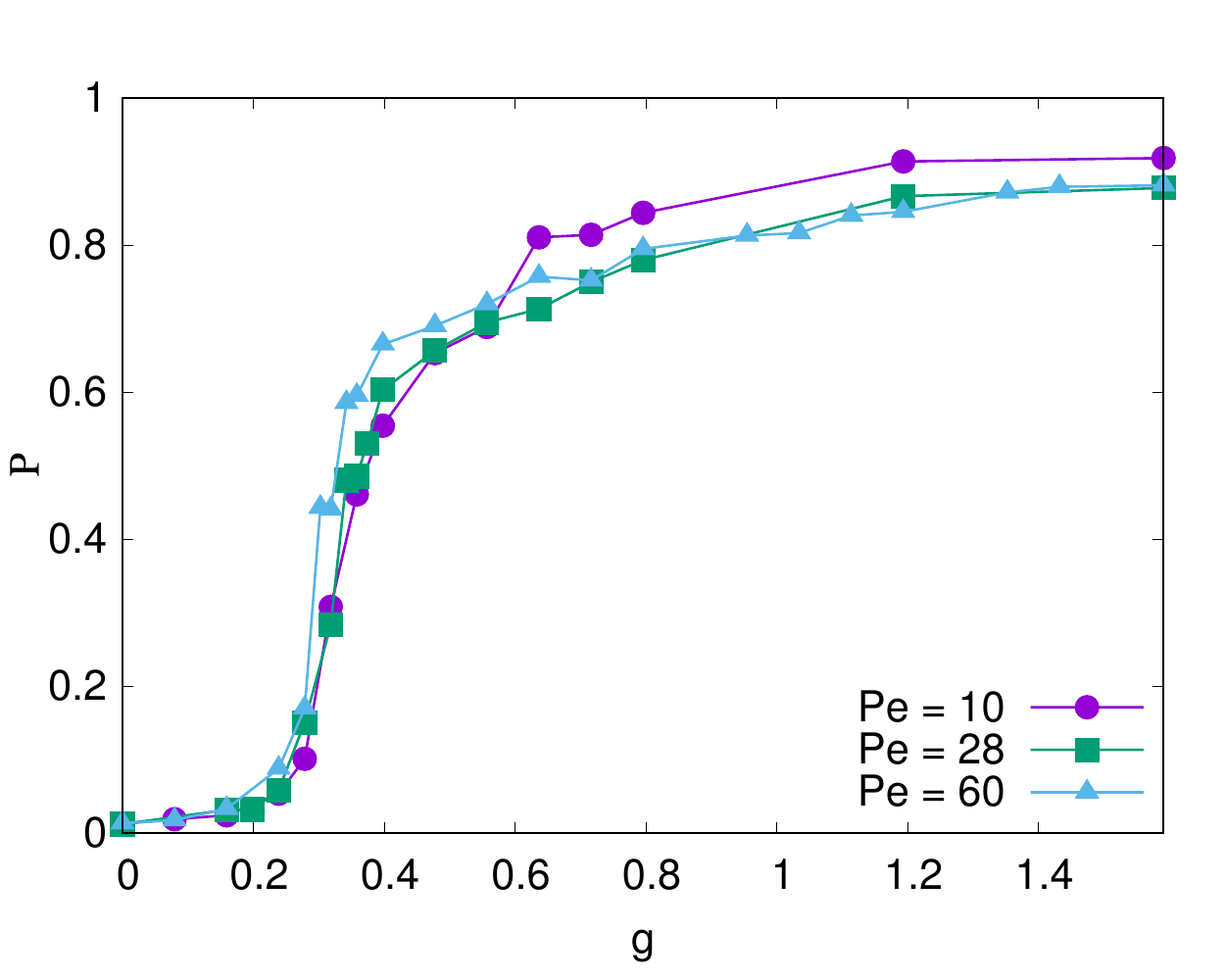}
\caption{Global polarization $P$ as a function of $g$ for different values of Pe   shown in the key. }
\label{fig:P}
\end{figure}

We start our study by  analyzing the onset of collective motion, or flocking transition, induced by the presence of velocity alignment interactions. 
As it is manifest from the model equation (\ref{eq:eomT}), the tendency of the particles to align their direction of motion competes with rotational noise. Thus, at small $g$, one expects noise to dominate, destroying the tendency to locally order. At higher $g$, the strength of the interaction might be large enough to overcome the random reorientation of the particles  and eventually lead to large-scale ordering in the space of velocities. 
Global order in the system can thus be characterized by the average polarization, $P =\langle  || \frac{1}{N}  \sum_{i} \textbf{n}_{i} ||\rangle$ (where $\langle * \rangle$ denotes an average over different steady-states). As shown in Fig. \ref{fig:P}, the system orders as we increase $g$, from where we can infer the emergence of directed motion (flocking): a macroscopic fraction of the system moves, in average, along a preferred direction. 
We identify the onset of flocking $g^c$ by $P(g=g_c) \gtrsim 0.35$. As shown in Fig. \ref{fig:P}, the onset of flocking seems, within our numerical accuracy, independent of the choice of Pe 
and equal to $g^{c} \approx 0.34$. 
This result is in agreement with previous particle-based simulation studies  \cite{Aitor} and with theoretical treatments that consider the coupling between velocity and density at a coarse-grained level \cite{Farrell, PeruaniChetrite}. In \cite{Farrell}, the authors use a mesoscopic description of excluded volume effects, encoded in a density-dependent velocity, and find that the onset of flocking in their effective theory is located at $g^{c} = \frac{2}{4 \pi \sigma^2 \rho} $, here, $g^c= 0.312$, independently of Pe.  This predicted value is reasonably close to our numerical estimate, although  we find it slightly above. As noted in  \cite{Farrell}, this is consistent with the fact that the theoretical prediction is based on a linear stability analysis which can only access the limit of stability of the disordered state or spinodal line. 


\section{Phase separation}
In the absence of alignment ($g=0$), particles slow down due to collisions, resulting in the decrease of particles' velocity with increasing local density. This creates a positive feed-back by which more particles accumulate in denser regions and thus, slow down,  eventually triggering a complete phase separation between a dense, slow-swimming phase and a dilute, fast-swimming one. This feed-back mechanisms leading to MIPS is usually described at a coarse-grained level by an effective density-dependent velocity $v(\rho)$, that decreases sufficiently fast with the density \cite{TailleurCates2008, CatesTailleur2013, Bialke2013, Stenhammar2013, CatesRev}. 


\begin{figure}
\onefigure[scale=0.65]{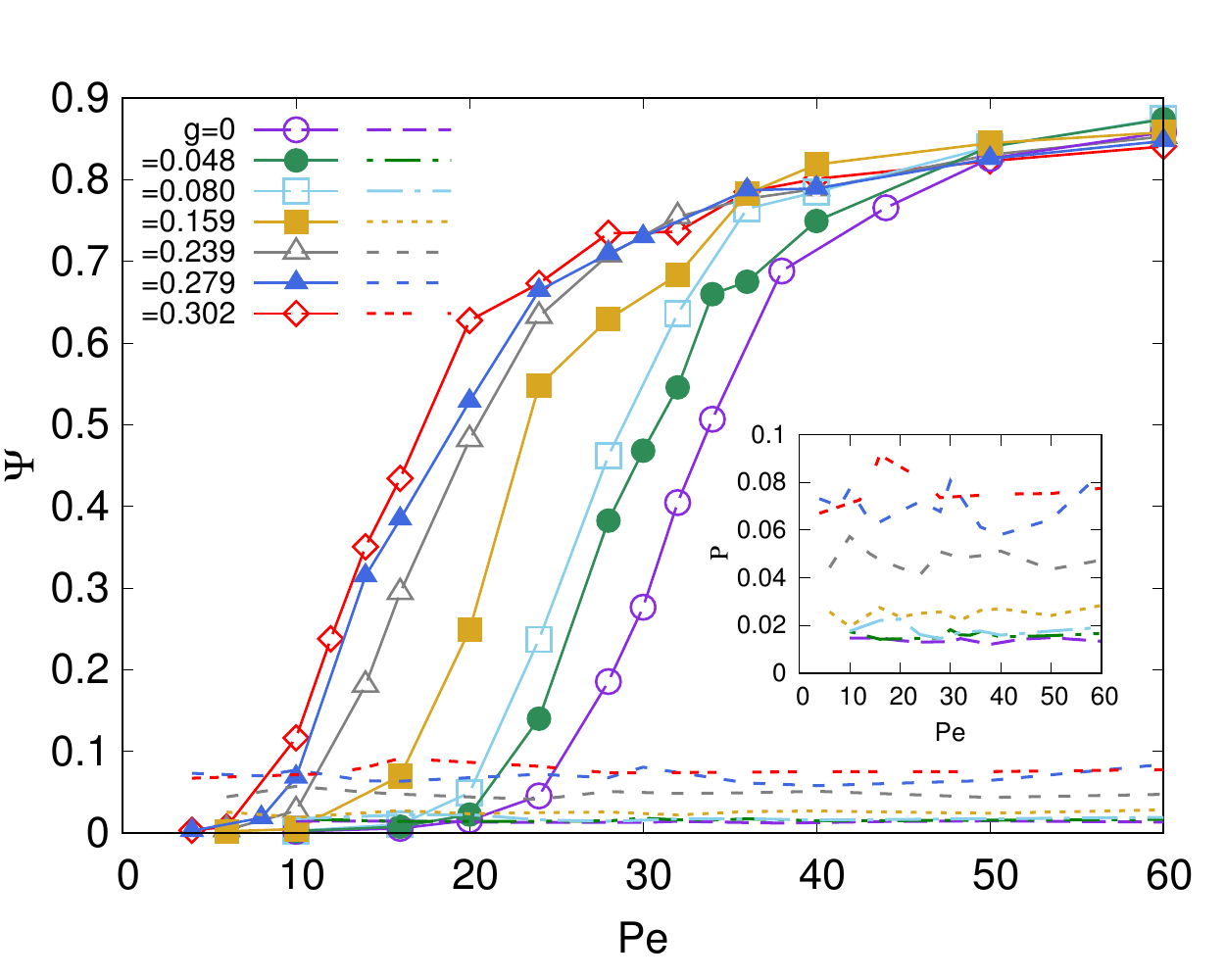}
\caption{Probability for a particle to belong to the largest cluster, $\Psi$, as a function of  $\text{Pe}$ for several values of $g$ shown in the key. Dashed lines show the global polarization $P$ of the system at each value of $\text{Pe}$ and $g$ (the inset shows the detailed view). Even though particles aggregate into clusters as $\text{Pe}$ increases, the system remains globally disordered, showing that phase separation is not induced by the emergence of flocking.}
\label{fig:2}
\end{figure}

We identify the onset of MIPS by the emergence of a large dense cluster (large in the sense that it represents a finite fraction of the total system). 
We define a cluster as a connected set of particles distant of less than  $1.4 \sigma$ (corresponding with the location of the first peak of the pair distribution function).
Once clusters have been identified, we can compute the fraction of particles belonging to the largest one, $\Psi$. We  thus use $\Psi$  as a phenomenological order parameter to identify the onset of  phase separation \cite{Bialke2013, Levis2017, Aitor}.
As shown in Fig. \ref{fig:2}, $\Psi$ becomes finite as Pe is increased. 
To systematically locate the onset of phase separation we consider the value of the $\text{Pe}$ for which the fraction of particles belonging to the macroscopic cluster is $\Psi = 0.35$. According to this criterion, in the absence of alignment ($g=0$),  MIPS occurs for  $\text{Pe}>\text{Pe} ^{*} = 31 \pm 2$.

We focus now on the impact of alignment on the aggregation of ABP reported above. 
We proceed by slowly varying the coupling parameter $g$ and computing the fraction of particles belonging to the largest cluster as a function of Pe, see Fig. \ref{fig:2}. 
Since the flocking phase transition is approached from below ($g<g^c$), the system does not develop any global net polarization in this regime. 
However, as shown in Fig. \ref{fig:2}, as the particle's tendency to align increases, the emergence of a large structure  takes place at lower values of Pe. This means that it is easier for particles to aggregate in the presence of  alignment. 
Is  this aggregation phenomena in the presence of velocity-alignement dominated by the feedback mechanism behind MIPS? Or is it rather controlled by the mechanisms responsible of structure formation in the form of traveling fronts in polar active matter?


\begin{figure}
\onefigure[scale=0.65]{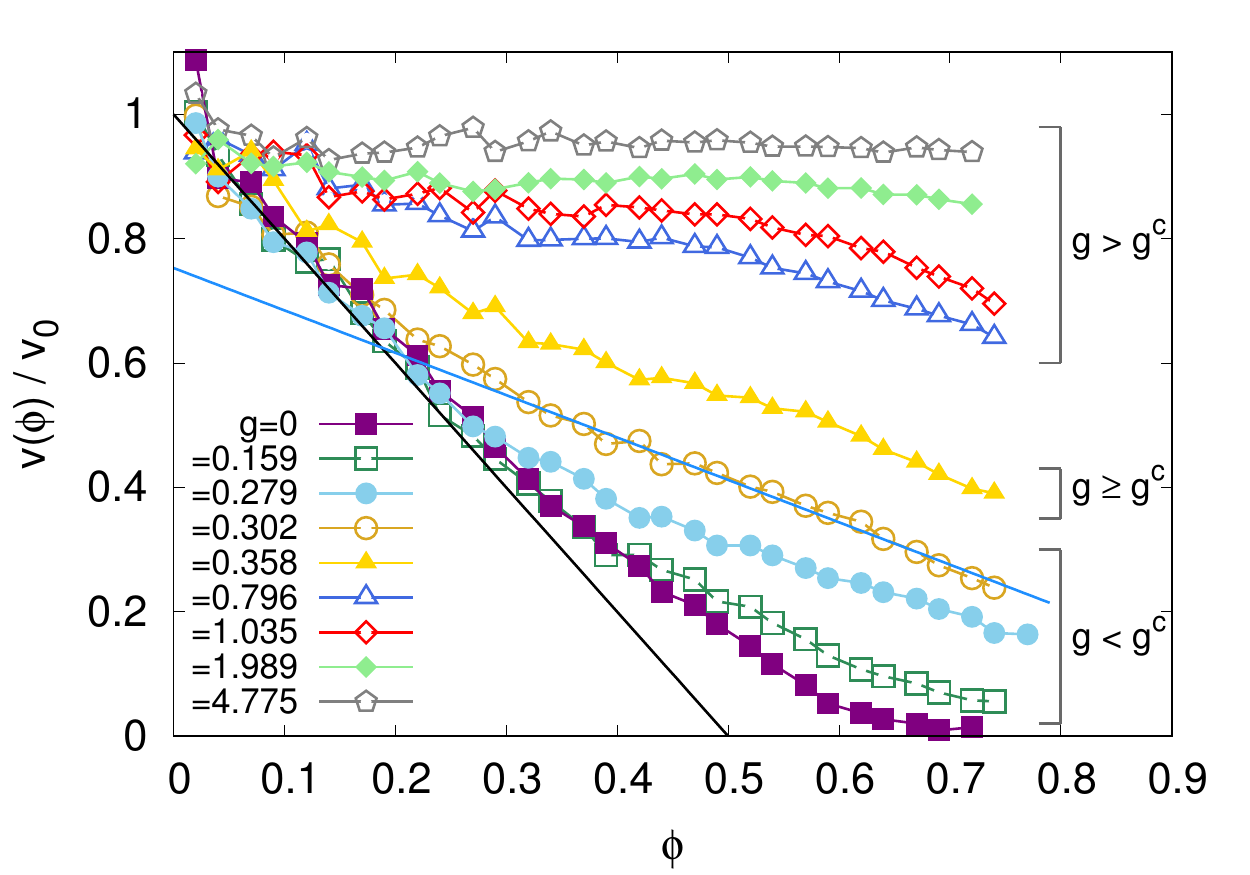}
\caption{Average velocity projected in the direction of self-propulsion as a function of the  packing fraction for different values of $g$ and $\text{Pe} = 60$. Below  flocking ($g\leq 0.302$), the velocity displays two different linear decays at low and high densities : $v(\phi)/  v_{0}=1 - 2 \phi$ (continuous black line) and  $v(\phi)/  v_{0}=0.75 - 0.68 \phi$ (continuous blue line), respectively.  Above flocking ($g\geq 0.796$) the velocity decreases slower than linearly (and eventually saturates at large coupling).}
\label{fig:7}
\end{figure}

In order to answer these questions and better understand the aggregation mechanism for $0<g<g^c$, we analyze the dependence of particles' velocity with local density. 
We compute the average velocity  projected in the direction of self-propulsion as a function of the density $v(\phi)$. We proceeded by  subdividing the simulation box in  cells and measure $v_{j} = \frac{1}{N_{{j}}} \big| \sum_{i \in c_{j}} \dot{\textbf{r}}_{i} \cdot \textbf{n}_{i} \big|$, where $N_{{j}}$ is the number of particles in cell $c_{j}$ and $\phi_j=\pi N_j/(4c_j^2)$. 
The data obtained after averaging over $10^4$ configurations is shown in Fig. \ref{fig:7}, where the local mean velocity in the self-propulsion direction, $v$, is plotted as a function of the local density. 

Below the onset of flocking,  $v(\phi)$ decreases  with increasing local density, indicating that particles are slowed down due to crowding effects, as expected from the MIPS scenario. 
However, above the flocking phase transition, directed motion emerges, such that a macroscopic (finite) fraction of the system travels coherently at the same mean velocity. In this regime the system is globally oriented and, thus, a macroscopic fraction of particles in the system do not block each other when they collide, but rather mutually align, giving rise to clusters that move coherently \cite{Aitor}. 
We  conclude that, below the flocking phase transition,  the particles' aggregation arises from the competition between self-propulsion and steric effects,  through an effective density-dependent velocity. Therefore, the kinetic self-trapping mechanism responsible of MIPS survives in systems of polar ABP below the onset of flocking and it is responsible for the emergence of large clusters (as shown in Fig. \ref{fig:2}). 
However, in contrast with  mesoscopic approaches that consider a given functional decay of $v(\rho)$  (usually exponential) \cite{Farrell, PeruaniChetrite}, it is evident from the data Fig. \ref{fig:7} that alignment alters the decay of $v(\phi)$. Indeed, the  linear decay of velocity with density of isotropic ABP \cite{Stenhammar2013} is maintained at finite $g$ for low enough densities, but shifts to a slower, yet linear, decay at higher $\phi$ (see discussion below).     

\begin{figure*}
\centering
\includegraphics[width=1.0\textwidth,height=0.37\textheight ,trim={0cm 3cm 0cm 0cm}]{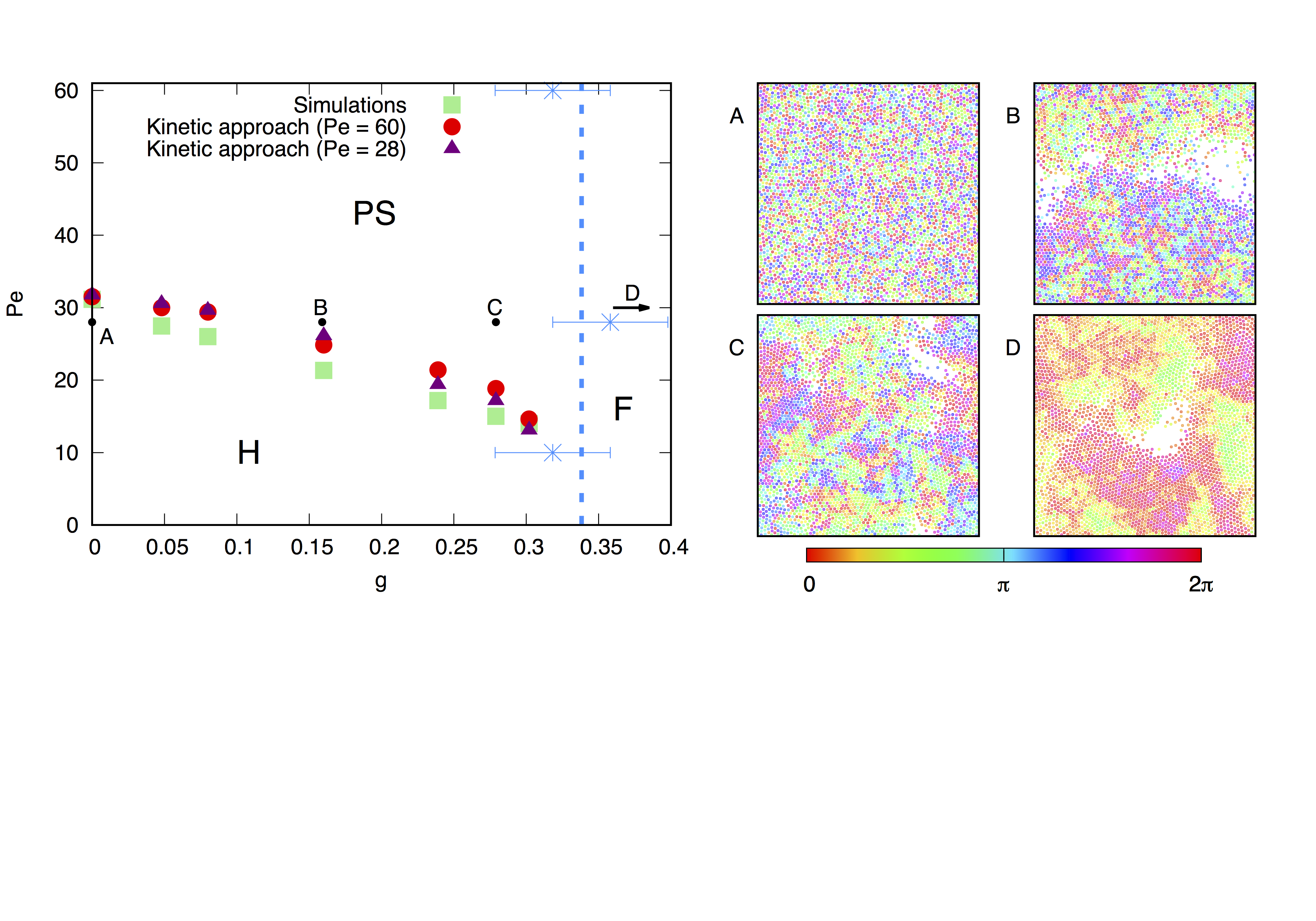}
\caption{Left: Phase diagram in the $(g,\,\text{Pe})$ plane. Squares mark the onset of MIPS obtained from simulation results. Circles and triangles show the transition obtained from the kinetic approach for different values of $\text{Pe}$. Asterisks indicate the onset of flocking at three different values of $\text{Pe}$ obtained from simulations. The onset of flocking is, within our numerical accuracy, independent of the self-propulsion velocity and at $g^{c} \approx 0.34$ (dashed vertical line). Right: Snapshots of the system at four different points of the phase diagram, A: (0,28); B: (0.159,28); C: (0.279,28); D: (0.796,28). The gradient of colors indicates the orientation of particles.}
\label{fig:6}
\end{figure*}

From the data shown in Fig. \ref{fig:2}, we extract the dependence of the MIPS threshold as a function of the alignment strength, denoted $\text{Pe}^*(g)$ and  shown by green symbols in the phase diagram Fig.  \ref{fig:6}. 
The vertical dashed line in the phase diagram shows the onset of flocking $g^c$. Above it, in the ordered region, the formation of clusters cannot be attributed to MIPS but to the directed motion of polar self-propelled particles (flocking phase denoted F). Below the onset of flocking, the system is controlled by the physics of ABP: at low self-propulsion the system is in a homogeneous gas phase (denoted H), while  for  $\text{Pe}>\text{Pe}^*(g)$ the system exhibits MIPS (phase separated region, denoted PS). The MIPS threshold  $\text{Pe}^*(g)$ decreases with $g$: alignment favors phase separation. 
As illustrated by the snapshots Fig.  \ref{fig:6}, at $g = 0$, the orientation of neighboring particles, represented by a color code, is completely decorrelated. As $g$ increases, approaching the flocking phase transition from below,  orientational correlations grow, as evidenced by the growing size of the (polarized) single-color domains of particles. The growth of such orientational correlations with $g$  is the crucial feature that explains why velocity-alignment facilitates MIPS.

\begin{figure*}
\centering
\includegraphics[width=1\textwidth,height=0.27\textheight ,trim={0cm 7cm 0cm 13cm}]{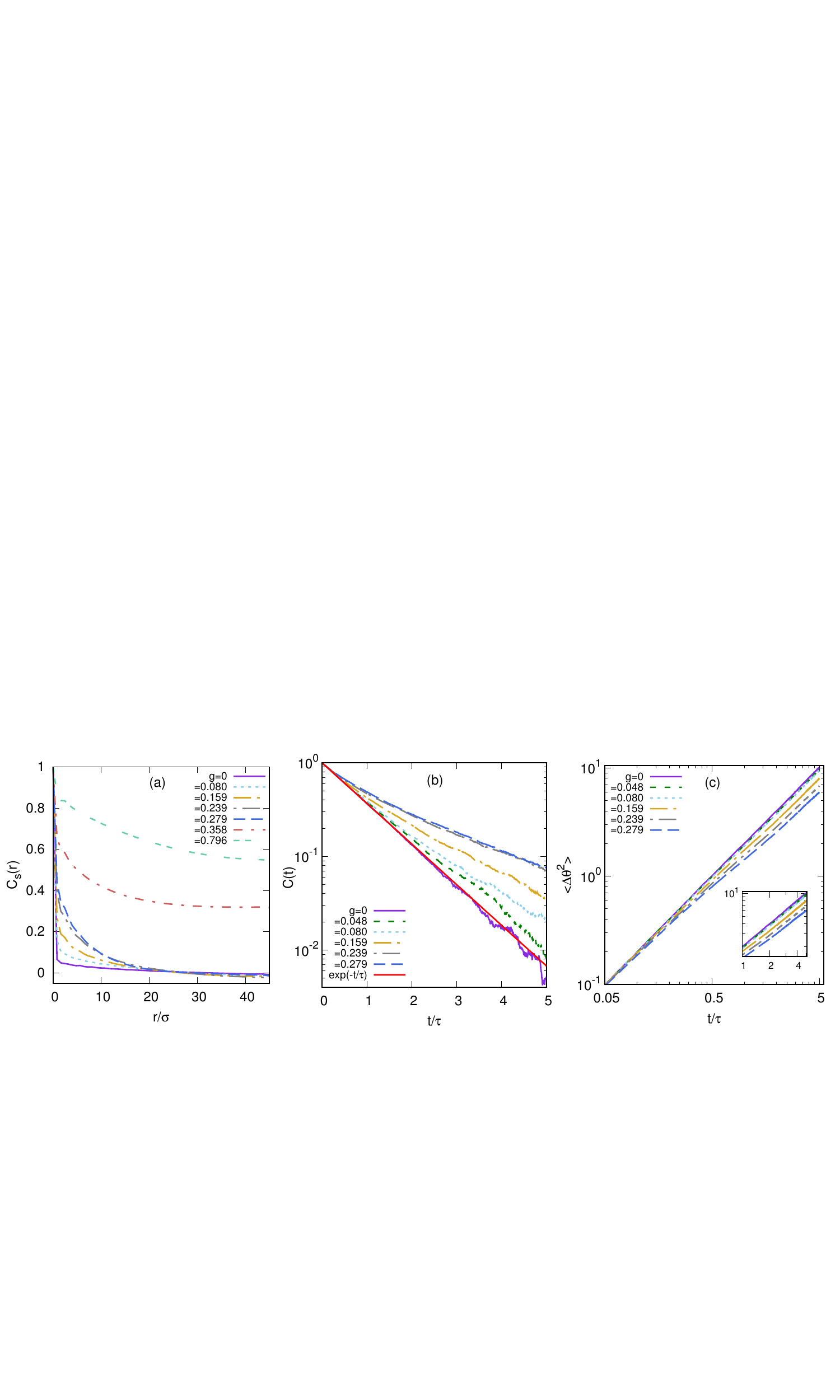}
\caption{(a): Orientational space-correlation function $C_s(r)$ for different values of $g$ and at constant $\text{Pe}=60$. (b): Log-linear plot of the orientational time-correlation function $C(t)$ for different values of $g$ and at constant $\text{Pe}=60$. The exponential decay $\exp(-t/\tau)$ is also shown for comparison. (c): Log-log plot of the average mean-square angular displacement for different values of $g$ and at constant $\text{Pe}=60$. }
\label{fig:8}
\end{figure*}

To quantify local orientational correlations we compute the correlation function 
\begin{equation}
C_{s}(r) = \langle\textbf{n}_{i} (t_s)\cdot \textbf{n}_{j} (t_s) \rangle_{|{\bold{r}_i-\bold{r}_j}|=r}
\end{equation}
in the steady state (i.e. $t_s\gg \tau$). 
In Fig. \ref{fig:8}(a) we show $C_{s}$ for several couplings across the onset of flocking. As expected, spatial correlations grow as $g$ increases, indicating the appearance of polarized domains. Above the onset of flocking, $C_s$ does not decay to zero, indicating that the system is globally ordered.  
To quantify the impact of the growth of spatial correlations on the orientational dynamics of particles, we also compute  orientational correlations in time, captured by the  self-correlation function
\begin{equation}
C(t) = \frac{1}{N}\sum_i\langle \textbf{n}_{i}(t_s) \cdot \textbf{n}_{i}(t+t_s)\rangle
\end{equation} 
As expected, time correlations decay exponentially and the relaxation becomes slower for  $g>0$ [see Fig. \ref{fig:8}(b)], meaning that the orientation of particles has a longer memory in the presence of alignment or, put in different words, their dynamics is more persistent and their random reorientation slower. 

From a kinetic point of view, the slowdown of the orientational dynamics due to the appearance of polarized domains should have an impact on the rate of absorption and evaporation of particles in clusters. Indeed, when two particles collide they get stuck until they reorient in a time scale that, as shown in Fig. \ref{fig:8}(b), is larger in the presence of alignment. 
Following  \cite{Redner2013}, the rate of absorption of particles in a cluster can be expressed as 
\begin{equation}
k_{\text{in}} = \frac{\rho_{l} v_{0}}{\pi}
\label{eq:11}
\end{equation}
where $\rho_{l}$ is the  number density of the low-density gas background, and the rate of evaporation 
\begin{equation}
k_{\text{out}} = \frac{\kappa \Da_{\theta}(g)}{\sigma}
\label{eq:12}
\end{equation}
where $\Da_{\theta}(g)$ is the effective rotational diffusion coefficient which, of course, generically depends on $g$. The parameter $\kappa$ quantifies the total number of particles that leave the cluster in each escape event.
The rate of evaporation of particles, $k_{\text{out}}$, depends on the time $\Da_{\theta}(g)^{-1}$ needed for a particle in the surface to randomly reorient, and pick an  orientation that allows it to move away from the cluster. Since the orientations are correlated for longer times as alignment strength increases, $k_{\text{out}}$ will decrease with $g$, while the rate of absorption $k_{\text{in}} $ remains unchanged. As a result, the aggregation of particles into clusters is enhanced, favoring the system's phase separation.

In order to account for the dependence of the onset of MIPS with $g$, we extend the kinetic approach given by Redner et al. \cite{Redner2013} to our case with velocity alignement.
In the steady-state,  the rate of absorption and  evaporation should be equal, leading to the gas  density
\begin{equation}
\rho_{l} = \frac{\kappa  \pi}{\sigma v_{0}}\Da_{\theta}(g)
\label{eq:17}
\end{equation}
The average  density of the system can be expressed as
\begin{equation}
\overline{\rho} = f \rho_{c} + (1 - f) \rho_{l} 
\label{eq:18}
\end{equation}
where $f$ is the fraction of particles belonging to the dense phase (i.e. the largest cluster)  and $\rho_{c}$  its number density.
We identify the onset of MIPS with the limit of coexistence $f = 0$, such that eq. \ref{eq:18} together with eq. \ref{eq:17} leads to an expression of the onset of MIPS as a function of the coupling strength:
\begin{equation}
\text{Pe}^{*}(g)=\frac{\kappa \pi }{\sigma^{2} \overline{\rho} D_{\theta}} \Da_{\theta}(g)
\label{eq:19}
\end{equation} 
All the $g$-dependence is encoded in the effective rotational diffusivity. Since, as already discussed, $\Da_{\theta}$ decreases with $g$, this simple argument allows us to rationalize why the onset of MIPS is shifted towards lower Pe in the presence of alignment.

In order to make a quantitative comparison between our simulation results and the prediction from the kinetic mean-field model, we use the onset of MIPS extracted from the simulations at $g=0$  as an input in eq. \ref{eq:19},  with  $\Da_{\theta} = D_{\theta}$,  to estimate $\kappa \approx 5$. 
We further assume  $\kappa$ to be independent of $g$ and measure $\Da_{\theta}$ from the long time behavior of the angular mean-square displacement
\begin{equation}
\langle \Delta\theta^2\rangle= \frac{1}{N}\sum_i \langle [\theta_{i}(t) - \theta_{i}(0)]^{2}\rangle \to_{t\to\infty}2\Da_{\theta}(g)t
\end{equation}
This quantity is plotted in Fig. \ref{fig:8} (c), where it is shown that  $\Da_{\theta}$ decreases with $g$, as expected from the previous discussion of $C(t)$. From the measurement of $\Da_{\theta}$, we  estimate the onset of MIPS, Pe$^*(g)$, using eq. \ref{eq:19}. Such estimation  is shown in the phase diagram Fig. \ref{fig:6} in red and purple symbols (for Pe$=60$ and 28, respectively). The results obtained from the kinetic model and the structural criterion on $\Psi$ (see Fig. \ref{fig:2}) are in considerable agreement and show the same overall behavior:  phase separation can be triggered by a mechanism analog to MIPS of apolar ABP in the presence of alignment. As illustrated in the sequence of snapshots Fig. \ref{fig:6}, increasing $g$ can drive MIPS at constant Pe. 

As shown in Fig. \ref{fig:7}, the decrease of $v(\phi)$  below the onset of flocking shows two clearly distinct regimes. In the dilute regime, the decay is independent of the alignment interaction between particles (as expected since in dilute regions particles are on average distant of more than the  alignment interaction range).
Conversely, at high packing fraction, the decay is slower in the presence of alignment, showing that the latter has an important impact on the dynamics of the system, even below the onset of flocking. Strikingly, the linear decay slope of $v(\phi)$ at high densities is, up to numerical accuracy, independent of the precise value of $g$. There are therefore two well-defined distinct regimes, indicating a change on the large-scale properties of the system as the density is varied. This suggests that such crossover is associated to the emergence of a MIPS, which,  in the presence of alignment, is characterized by a dense phase with enhanced orientational correlations.    


\section{Conclusion and Discussion}
We find that polar ABP subjected to local alignment interactions undergo a MIPS-like phase separation, in that  the aggregation mechanism below the onset of polar order is controlled by the competition between self-propulsion and steric effects. Moreover, the formation of non-polar clusters is favored when increasing the alignment interaction, thus fostering the complete phase separation of the system. Such behavior is understood as arising from the enhancement of orientational correlations which largely suppresses the random reorientation of the particles. 
The microscopic character of our approach allows to explore the kinetic mechanisms responsible of clustering in systems of ABP with velocity alignment  and critically asses  the appropriateness of the MIPS picture in this context. The understanding  of the two different regimes in the decay of the velocity we identified from a coarse-grained theory calls for further developments, since the mesoscopic descriptions in terms of a single-mode density-dependent velocity do not correctly capture the competition between self-propulsion and excluded volume in the presence of alignment.

\acknowledgments
D.L. acknowledges received funding from the European Union's Horizon 2020 research and innovation programme under the Marie Sklodowska-Curie (IF) grant agreement No 657517. I.P. acknowledges MINECO and DURSI for financial support under projects FIS2015-67837-P and 2017SGR-884, respectively.
All the authors acknowledge support from the COST Action MP1305-Flowing Matter.

\bibliographystyle{eplbib}
\bibliography{elena-mips}

\begin{thebibliography}{10}
\expandafter\ifx\csname url\endcsname\relax\def\url#1{\texttt{#1}}\fi

\bibitem{RamaswamyRev}
\Name{Ramaswamy S.} \REVIEW{Annu. Rev. Cond. Matt. Phys.}{1}{2010}{323}.

\bibitem{VicsekRev}
\Name{Vicsek T. \and Zafeiris A.} \REVIEW{Phys. Rep.}{517}{2012}{71}.

\bibitem{MarchettiRev}
\Name{Marchetti M.~C., Joanny J.-F., Ramaswamy S., Liverpool T.~B., Prost J.,
  Rao M. \and Simha R.~A.} \REVIEW{Rev. Mod. Phys.}{85}{2013}{1143}.

\bibitem{Zhang2010}
\Name{Zhang H.-P., Beer A., Florin E.-L. \and Swinney H.~L.} \REVIEW{Proc. Nat.
  Ac. Sci. USA}{107}{2010}{13626}.

\bibitem{Peruani2012}
\Name{Peruani F., Starru{\ss} J., Jakovljevic V., S{\o}gaard-Andersen L.,
  Deutsch A. \and B{\"a}r M.} \REVIEW{Phys. Rev. Lett.}{108}{2012}{098102}.

\bibitem{Chate2017bac}
\Name{Nishiguchi D., Nagai K.~H., Chat{\'e} H. \and Sano M.} \REVIEW{Phys. Rev.
  E}{95}{2017}{020601}.

\bibitem{Theurkauff2012}
\Name{Theurkauff I., Cottin-Bizonne C., Palacci J., Ybert C. \and Bocquet L.}
  \REVIEW{Phys. Rev. Lett.}{108}{2012}{268303}.

\bibitem{Palacci2013}
\Name{Palacci J., Sacanna S., Steinberg A.~P., Pine D.~J. \and Chaikin P.~M.}
  \REVIEW{Science}{}{2013}{1230020}.

\bibitem{Buttinoni2013}
\Name{Buttinoni I., Bialk{\'e} J., K{\"u}mmel F., L{\"o}wen H., Bechinger C.
  \and Speck T.} \REVIEW{Phys. Rev. Lett.}{110}{2013}{238301}.

\bibitem{BechingerRev}
\Name{Bechinger C., Di~Leonardo R., L{\"o}wen H., Reichhardt C., Volpe G. \and
  Volpe G.} \REVIEW{Rev. Mod. Phys.}{88}{2016}{045006}.

\bibitem{Ginot2015}
\Name{Ginot F., Theurkauff I., Levis D., Ybert C., Bocquet L., Berthier L. \and
  Cottin-Bizonne C.} \REVIEW{Phys. Rev. X}{5}{2015}{011004}.

\bibitem{Ginot2018}
\Name{Ginot F., Theurkauff I., Detcheverry F., Ybert C. \and Cottin-Bizonne C.}
  \REVIEW{Nat. Comm.}{9}{2018}{696}.

\bibitem{Schaller2010}
\Name{Schaller V., Weber C., Semmrich C., Frey E. \and Bausch A.~R.}
  \REVIEW{Nature}{467}{2010}{73}.

\bibitem{Huber2018}
\Name{Huber L., Suzuki R., Kr{\"u}ger T., Frey E. \and Bausch A.~R.}
  \REVIEW{Science}{}{2018}{eaao5434}.

\bibitem{Deseigne2010}
\Name{Deseigne J., Dauchot O. \and Chat{\'e} H.} \REVIEW{Phys. Rev.
  Lett.}{105}{2010}{098001}.

\bibitem{Bricard2013}
\Name{Bricard A., Caussin J.-B., Desreumaux N., Dauchot O. \and Bartolo D.}
  \REVIEW{Nature}{503}{2013}{95}.

\bibitem{Romanczuk2012active}
\Name{Romanczuk P., B{\"a}r M., Ebeling W., Lindner B. \and Schimansky-Geier
  L.} \REVIEW{Eur. Phys. J. Special Topics}{202}{2012}{1}.

\bibitem{Fily2012}
\Name{Fily Y. \and Marchetti M.~C.} \REVIEW{Phys. Rev.
  Lett.}{108}{2012}{235702}.

\bibitem{Redner2013}
\Name{Redner G.~S., Hagan M.~F. \and Baskaran A.} \REVIEW{Phys. Rev.
  Lett.}{110}{2013}{055701}.

\bibitem{CatesTailleur2013}
\Name{Cates M. \and Tailleur J.} \REVIEW{EPL}{101}{2013}{20010}.

\bibitem{Bialke2013}
\Name{Bialk{\'e} J., L{\"o}wen H. \and Speck T.}
  \REVIEW{EPL}{103}{2013}{30008}.

\bibitem{Stenhammar2014}
\Name{Stenhammar J., Marenduzzo D., Allen R.~J. \and Cates M.~E.} \REVIEW{Soft
  Matter}{10}{2014}{1489}.

\bibitem{Levis2017}
\Name{Levis D., Codina J. \and Pagonabarraga I.} \REVIEW{Soft
  Matter}{13}{2017}{8113}.

\bibitem{Digregorio2018}
\Name{Digregorio P., Levis D., Suma A., Cugliandolo L.~F., Gonnella G. \and
  Pagonabarraga I.} \REVIEW{arXiv preprint arXiv:1805.12484}{}{2018}{}.

\bibitem{Vicsek1995}
\Name{Vicsek T., Czir{\'o}k A., Ben-Jacob E., Cohen I. \and Shochet O.}
  \REVIEW{Phys. Rev. Lett.}{75}{1995}{1226}.

\bibitem{Toner1995}
\Name{Toner J. \and Tu Y.} \REVIEW{Phys. Rev. Lett.}{75}{1995}{4326}.

\bibitem{Gregoire2004}
\Name{Gr{\'e}goire G. \and Chat{\'e} H.} \REVIEW{Phys. Rev.
  Lett.}{92}{2004}{025702}.

\bibitem{Chate2008}
\Name{Chat{\'e} H., Ginelli F., Gr{\'e}goire G. \and Raynaud F.} \REVIEW{Phys.
  Rev. E}{77}{2008}{046113}.

\bibitem{Chate2008Rev}
\Name{Chat{\'e} H., Ginelli F., Gr{\'e}goire G., Peruani F. \and Raynaud F.}
  \REVIEW{Eur. Phys. J. B}{64}{2008}{451}.

\bibitem{Mishra2010}
\Name{Mishra S., Baskaran A. \and Marchetti M.~C.} \REVIEW{Phys. Rev.
  E}{81}{2010}{061916}.

\bibitem{Solon2015}
\Name{Solon A.~P., Chat{\'e} H. \and Tailleur J.} \REVIEW{Phys. Rev.
  Lett.}{114}{2015}{068101}.

\bibitem{CatesRev}
\Name{Cates M.~E. \and Tailleur J.} \REVIEW{Annu. Rev. Cond. Matt.
  Phys.}{6}{2015}{219}.

\bibitem{TailleurCates2008}
\Name{Tailleur J. \and Cates M.~E.} \REVIEW{Phys. Rev.
  Lett.}{100}{2008}{218103}.

\bibitem{Szabo2006}
\Name{Szabo B., Sz{\"o}ll{\"o}si G.~J., G{\"o}nci B., Jur{\'a}nyi Z., Selmeczi
  D. \and Vicsek T.} \REVIEW{Phys. Rev. E}{74}{2006}{061908}.

\bibitem{Henkes2011}
\Name{Henkes S., Fily Y. \and Marchetti M.~C.} \REVIEW{Phys. Rev.
  E}{84}{2011}{040301}.

\bibitem{Peruani2011traffic}
\Name{Peruani F., Klauss T., Deutsch A. \and Voss-Boehme A.} \REVIEW{Phys. Rev.
  Lett.}{106}{2011}{128101}.

\bibitem{Farrell}
\Name{Farrell F. D.~C., Marchetti M.~C., Marenduzzo D. \and Tailleur J.}
  \REVIEW{Phys. Rev. Lett.}{108}{2012}{248101}.

\bibitem{Lam2015}
\Name{Lam K. D. N.~T., Schindler M. \and Dauchot O.} \REVIEW{New J.
  Phys.}{17}{2015}{113056}.

\bibitem{PeruaniChetrite}
\Name{Barr{\'e} J., Ch{\'e}trite R., Muratori M. \and Peruani F.} \REVIEW{J.
  Stat. Phys.}{158}{2015}{589}.

\bibitem{Aitor}
\Name{Mart{\'\i}n-G{\'o}mez A., Levis D., D{\'\i}az-Guilera A. \and
  Pagonabarraga I.} \REVIEW{Soft Matter}{}{2018}{2610}.

\bibitem{Tailleur2008}
\Name{Tailleur J. \and Cates M.} \REVIEW{Phys. Rev. Lett.}{100}{2008}{218103}.

\bibitem{SharmaBrader}
\Name{Sharma A. \and Brader J.~M.} \REVIEW{J. Chem. Phys.}{}{2016}{161101}.

\bibitem{Levis2018JCMP}
\Name{Levis D. \and Liebchen B.} \REVIEW{J. Phys. Condens. Matter}{}{2018}{}.

\bibitem{Peruani2008mean}
\Name{Peruani F., Deutsch A. \and B{\"a}r M.} \REVIEW{Eur. Phys.
  J.}{157}{2008}{111}.

\bibitem{Grossmann2012}
\Name{Grossmann R., Schimansky-Geier L. \and Romanczuk P.} \REVIEW{New J.
  Phys.}{14}{2012}{073033}.

\bibitem{Alarcon2013}
\Name{Alarc{\'o}n F. \and Pagonabarraga I.} \REVIEW{J. Mol.
  Liq.}{185}{2013}{56}.

\bibitem{Stenhammar2013}
\Name{Stenhammar J., Tiribocchi A., Allen R.~J., Marenduzzo D. \and Cates
  M.~E.} \REVIEW{Phys. Rev. Lett.}{111}{2013}{145702}.

\end{thebibliography}

\end{document}